\documentclass{ws-procs975x65}
\usepackage{graphicx}
\usepackage{latexsym}
\usepackage{caption}
\usepackage{amssymb}
\begin{document}
\newcommand{\ti}[1]{\mbox{\tiny{#1}}}
\newcommand{\im}{\mathop{\mathrm{Im}}}
\newcommand{\ap}[1]{\ ^{\mbox{\tiny{(#1)}}}\! }
\title{Circular motion in Reissner-Nordstr\"om spacetime}
\author{Daniela Pugliese$^{1}$, Hernando Quevedo$^{1,2,3}$ and Remo Ruffini$^{1, 2}$}
\address{
$^1$ ICRA—International Center for Relativistic Astrophysics,
Physics Department (G9), University of Rome, ``La Sapienza'',
Piazzale Aldo Moro 5, 00185 Rome, Italy
\\
E-Mail: Pugliese@icra.it\\
$^{2}$ ICRANet - C.C. Pescara, Piazza della
Repubblica 10, I-65100, Pescara, Italy,\\
E-Mail: Ruffini@icra.it\\
$^{3}$ Instituto de Ciencias Nucleares, Universidad Nacional
Aut$\acute{o}$noma de M$\acute{e}$xico
\\
E-Mail: quevedo@nucleares.unam.mx
}
\begin{abstract}
We study the motion of neutral test particles along circular orbits
in the Reissner-Nordstr\"om spacetime. We use the method of the
effective potential with the constants of motion associated to the
underlying Killing symmetries. A comparison between the black hole
and naked singularity cases is performed. In particular we find that
in the naked singularity case for $r<Q^2/M$ no circular orbits can
exist, this radius plays a fundamental role in the physics of the
naked singularity. For $r>Q^2/M$ and $1<Q/M<\sqrt{9/8}$ there are
two stability regions together with a region where all the circular
orbits are instable  and a zone in which all trajectory are possible
but not circular orbits, for $\sqrt{9/8}<Q/M<\sqrt{5}/2$ one
stability region and a region where all circular orbits are instable
appear, finally for $Q/M>\sqrt{5}/2$ there are all stable circular
orbits for $r>Q^2/M$.
%
\end{abstract}
\section{Introduction}
We consider the background of a Reissner-Nordstr\"om spacetime
described in standard spherical coordinates by the line element
\begin{equation}\label{11metrica}
ds^2=-\frac{\Delta}{r^2}dt^2+\frac{r^2}{\Delta}dr^2
+r^2\left(d\theta^2+\sin^2\theta d\phi^2\right)\ ,
\end{equation}
where $\Delta=r^2-2Mr +Q^2$; the horizon radii are given by
$r_{\pm}=M\pm\sqrt{M^2-Q^2}$. The associated electromagnetic
potential and field are respectively $ A=(Q/r)dt$ and  $
F=dA=-(Q/r^2)dt\wedge dr$.
\section{The effective potential for neutral test particles}
Metric (\ref{11metrica}) admits  the static Killing field
$\xi_{t}=\partial_{t}$ and the rotational Killing field
$\xi_{\phi}=\partial_{\phi} $. If
 $p^{a}\equiv\mu u^{a}$ is the four-vector energy momentum
of a neutral  particle of mass $\mu$, from the Killing vectors and
the geodetic motion we obtain two constants of motion related to the
Killing vectors, the timelike  Killing vector $\xi_{t}$ at infinity
related to the stationarity of the metric  and the spacelike Killing
vector $\xi_{\phi}$ related to the  axial symmetry: $E\equiv
-g_{\alpha\beta}\xi_{t}^{\alpha} p^{\beta}
=\frac{\Delta}{r^2}\dot{t}\mu$ is interpreted  for time-like
geodesics as representing the total energy of the particle as
measured by  a static observer at infinity, and   $ L \equiv
g_{\alpha\beta}\xi_{\phi}^{\alpha}p^{\beta} = r^2\dot{\phi}\mu$ is
interpreted as the angular momentum  of the particle. From the
magnitude $g_{\alpha\beta}p^{\alpha}p^{\beta}=-\mu^2$,  we find the
effective potential for a neutral particle in the background
(\ref{11metrica}) \cite{RuRR}
\begin{equation}\label{peff}
V\equiv\sqrt{\left(1+\frac{L^2}{\mu^2r^2}\right)\left(1+\frac{Q^2}{r^2}-\frac{2M}{r}\right)}
\ .
\end{equation}
The energy $E/\mu$ and the angular momentum $L/(\mu M)$ of a
(time-like) particle in a circular orbit of radius $r$ are obtained
from the condition $\partial V/\partial r=0$ as
\begin{equation}\label{df}
\frac{E}{\mu}=\frac{\Delta}{r\sqrt{r^2-3Mr+2Q^2}},\quad
\frac{L_{\pm}}{M\mu} =\pm
\frac{r\sqrt{r-\frac{Q^2}{M}}}{\sqrt{M}\sqrt{r^2-3Mr+2Q^2}}\ .
\end{equation}
\subsection{Black holes}
\begin{figure}[h!]
\centering
\begin{tabular}{c}
\includegraphics[scale=.9]{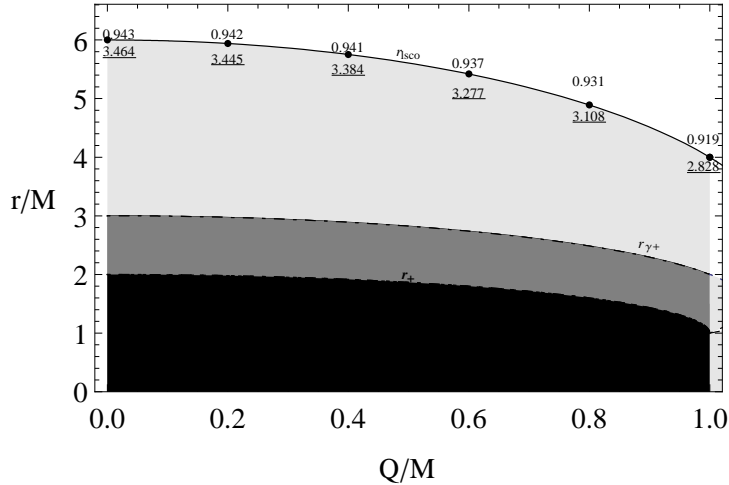}
\end{tabular}
\caption[font={footnotesize,it}]{The minimum radius $r_{lsco}/M$
(solid curve) is plotted as a function of the ratio $Q/M$ for the
case of black holes, $r_{+}=M+\sqrt{M^2-Q^2}$ is the outer horizon
while the dot-dashed curve is
$r_{\gamma^+}\equiv[3M+\sqrt{(9M^2-8Q^2)}]/2$. Black regions are
forbidden, no trajectories are possible there. In the region inside
the curves $r_{+}$ and $r_{\gamma^+}$ (gray region) no circular
orbits are possible. In the  region $r_{\gamma^+}$ and $r_{lsco}$,
(light-gray region) there are instable circular orbits. Finally for
$r>r_{lsco}$  any point represents a stable circular orbit. Numbers
close to the selected points represent the value of the energy
$E/\mu$ and the angular momentum $L/(\mu M)$ (underlined numbers) of
the last stable circular orbits.} \label{PcEqstab1ma}
\end{figure}

In the black hole case $(Q\leq M)$ circular orbits exist for
$r>r_{\gamma^+}\equiv[3M+\sqrt{(9M^2-8Q^2)}]/2$, where
$r=r_{\gamma^+}$ represents a photon orbit. The  study of the last
stable circular orbit ($r_{lsco}$) is sketched in
\fref{PcEqstab1ma}.
In the limiting Schwarzschild case with $Q=0$, we find $r=6M$, as
expected. In general, stable orbits are possible only for
$r>r_{lsco}$, whereas in the interval $r_{\gamma^+}<r<r_{lsco}$
there are only unstable orbits. Moreover, from \fref{PcEqstab1ma} we
conclude that as the charge-mass ratio increases from the
Schwarzschild value to the extreme black hole value $(Q=M)$, the
radius of the orbit and the corresponding energy and angular
momentum of the particle decrease\cite{Mio1Que,RuRR,Chandra,062}.
\subsection{Naked singularities}
\begin{figure}[h!]
\centering
\begin{tabular}{cc}
\includegraphics[scale=.71]{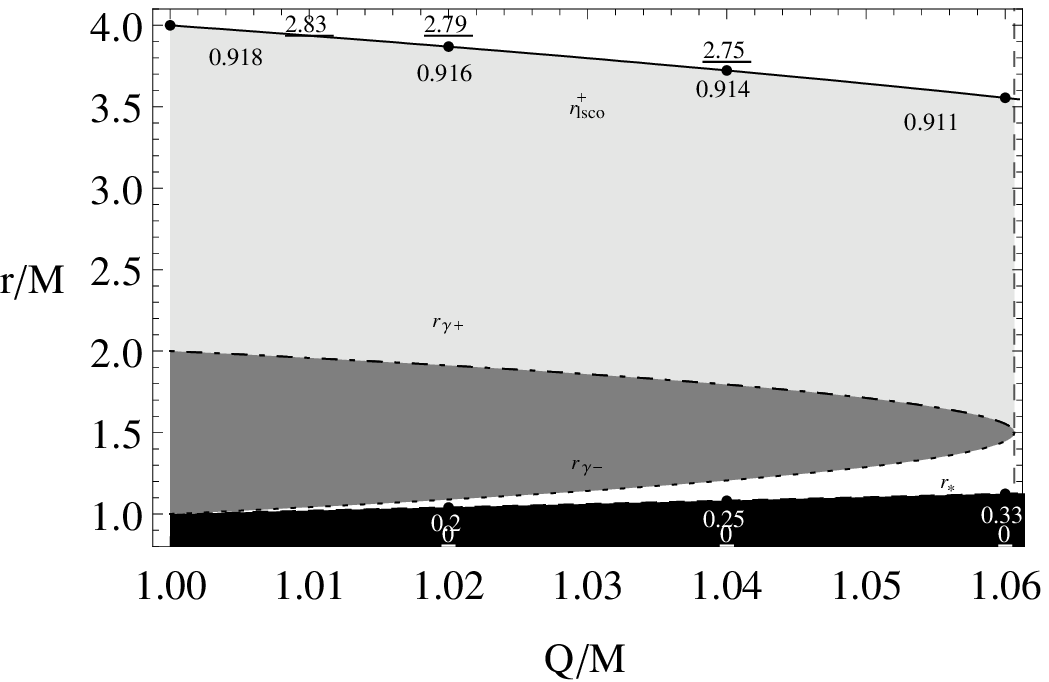}
\includegraphics[scale=.71]{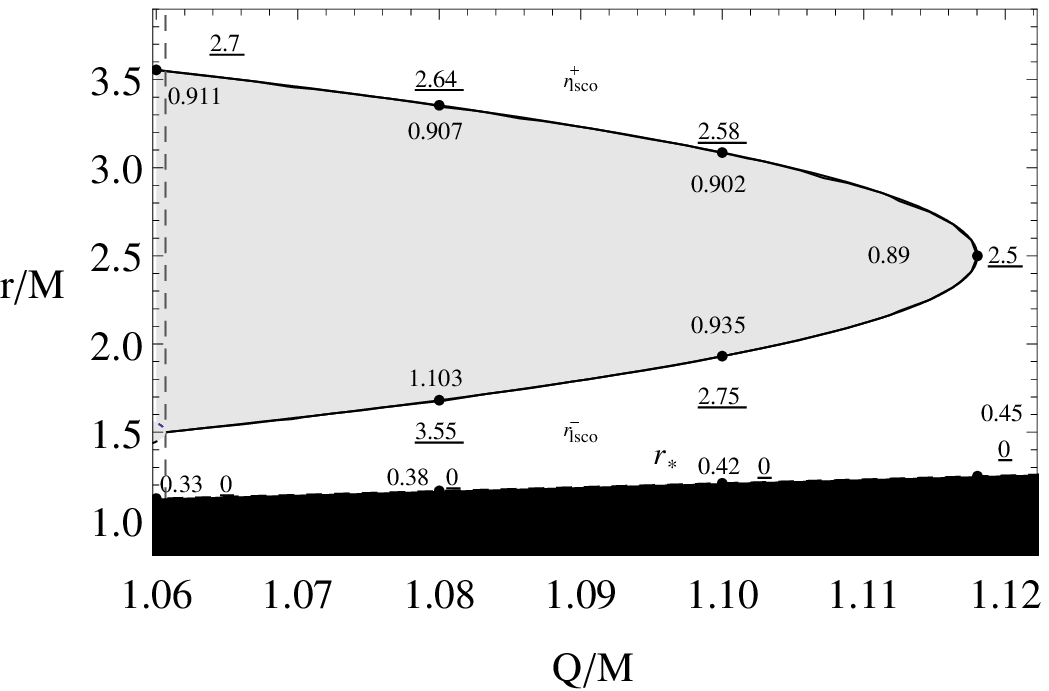}\\(a)&(b)\\
\includegraphics[scale=1]{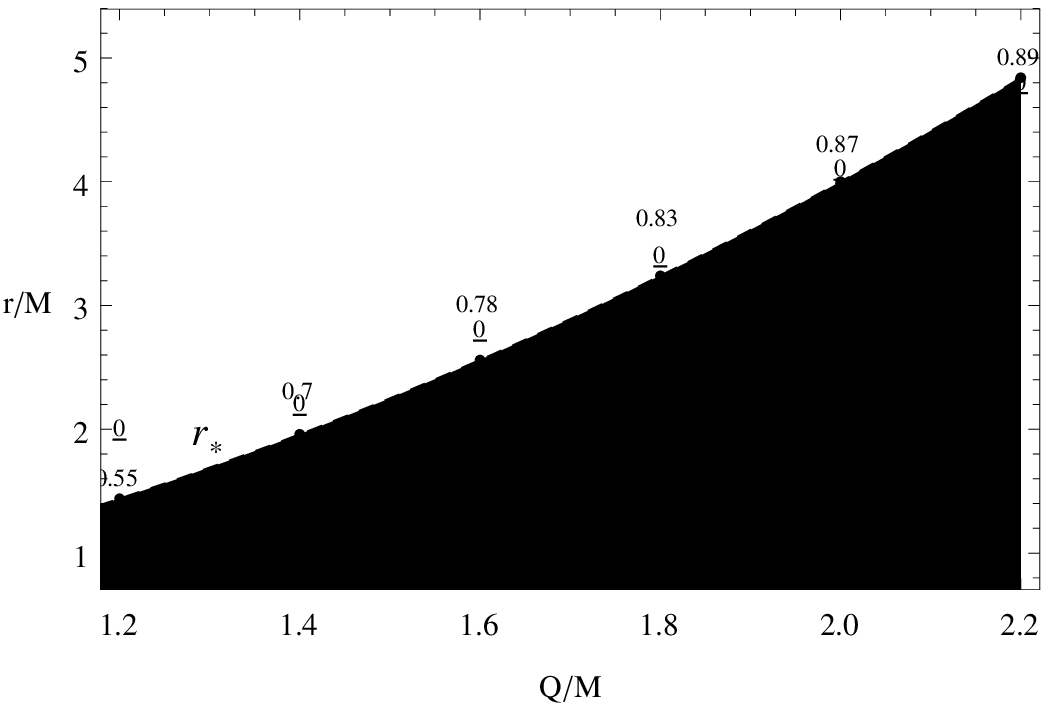}\\(c)
\end{tabular}
\caption[font={footnotesize,it}]{\footnotesize{The minimum radius
$r^{\pm}_{lsco}/M$ (solid curve) is plotted as a function of the
ratio $Q/M$ for the  naked singularities. The dashed curve
represents the classical radius $r_{*}= Q^2/M$ line, the dotted one
is $r_{\gamma^-}\equiv[3M-\sqrt{(9M^2-8Q^2)}]/2$, while the
dot-dashed curve is $r_{\gamma^+}\equiv[3M+\sqrt{(9M^2-8Q^2)}]/2$.
Black regions are forbidden no trajectories are possible, in the
gray regions all trajectories are possible but  circular orbits, in
the  light-gray regions there are instable circular orbits and
finally in the white regions there are all stable circular orbits.
Numbers close to the selected points represent the value of the
energy $E/\mu$ and the angular momentum $L/(\mu M)$ (underlined
numbers) of the last stable circular orbits. In \emph{(a)} region
$1<Q/M<\sqrt{9/8}$ is explored. In the region inside the curves
$r^+_{lsco}$ and $r_{\gamma^+}$  (lightgray region)  there are
instable circular orbits, inside $r_{\gamma^+}$ and $r_{\gamma^-}$
(Gray region) all trajectories are possible but circular orbits. In
$r>r^+_{lsco}$ any point represents a stable circular orbit. In
\emph{(b)} region $\sqrt{9/8}<Q/M<\sqrt{5}/2$ is explored. In the
region inside the curves $r^-_{lsco}$ and $r^+_{lsco}$ (lightgray
region) there are instable circular orbits, for $Q^2/M<r<r^-_{lsco}$
and $r>r^+_{lsco}$  any point represents a stable circular orbit.
Finally in \emph{(c)} region $Q/M\geq\sqrt{5}/2$ is explored, for
$r>r^*$ any point represents a stable circular orbit.} }
\label{PcEqstab1m}
\end{figure}

\begin{figure}[h!]
\centering
\begin{tabular}{c}
\includegraphics[scale=1.4]{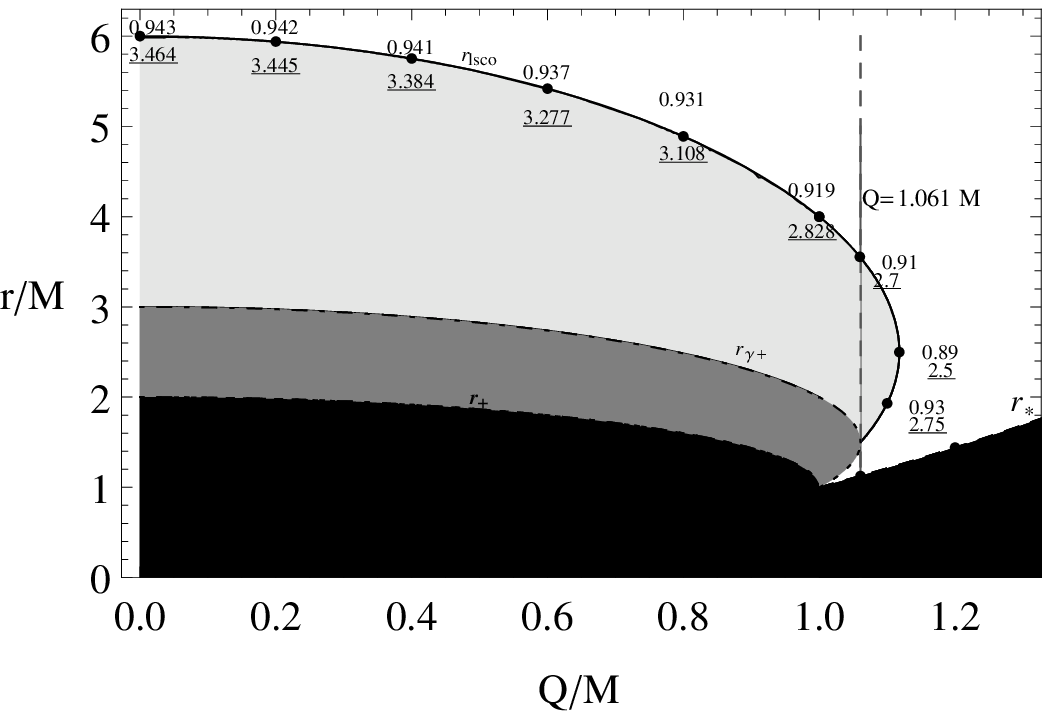}
\end{tabular}
\caption[font={footnotesize,it}]{\footnotesize{The minimum radius
$r^{\pm}_{lsco}/M$ (solid curve) is plotted as a function of the
ratio $Q/M$ in the range $[0,1.2]$. The classical radius $r_{*}=
Q^2/M$, radii $r_{\gamma^-}\equiv[3M-\sqrt{(9M^2-8Q^2)}]/2$,
$r_{\gamma^+}\equiv[3M+\sqrt{(9M^2-8Q^2)}]/2$ and the outer horizons
$r_{+}=M+\sqrt{M^2-Q^2}$ are also plotted. Black regions are
forbidden no trajectories are possible, in the  gray regions all
trajectories are possible but  circular orbits, in the light-gray
regions there are instable circular orbits
 and finally in the white regions there are all
stable circular orbits. Numbers close to the selected points
represent the value of the energy $E/\mu$ and the angular momentum
$L/(\mu M)$ (underlined numbers) of the last stable circular
orbits.} } \label{PcEqstab1mtree}
\end{figure}

In the naked singularity case $(Q>M)$ circular orbits are possible
only for $r\geq r_{*}$, where $ r_{*}=Q^2/M $ corresponds to the
classical radius of a particle with charge $Q$ and mass $M$. For
$r=r_{*}$ a time-like ``orbit'' with $L(r_{*})=0$ (particle at
``rest'') and $E/\mu=\sqrt{1-M^2/Q^2}$ is allowed. For a charge-mass
ratio in the interval $M<Q<\sqrt{9/8}M$ there exist time-like
circular orbits in the regions $r_{*}< r< r_{\gamma^-}$     and
$r>r_{\gamma^+}$, where
$r_{\gamma^{\pm}}\equiv[3M\pm\sqrt{(9M^2-8Q^2)}]/2$. The boundaries
$r=r_{\gamma^{\pm}}$ correspond to photon orbits, whereas in the
region $r_{\gamma^-}< r< r_{\gamma^+}$ only \emph{space-like}
circular orbits can exist. For $Q=\sqrt{9/8}M$ time-like circular
orbits exist for all $r> r_{*}$, except at the radius $r=(3/2)M$
which corresponds to a photon circular orbit \fref{PcEqstab1m}(a).
Finally, for $Q>\sqrt{9/8}M$ time-like circular orbits can exist for
all $r> r_{*}$. As for the stability of the circular orbits, we
found that for $M<Q<\sqrt{9/8}M$ there exist stable orbits only in
the regions $r_{*}< r< r_{\gamma^-}$  and $r>r^+_{lsco}$, and
unstable orbits are contained in $r_{\gamma^+}<r<r^+_{lsco}$. For
$Q=\sqrt{9/8}M$ all orbits are stable in the region $r>r_{*}$,
except at $r=(3/2)M$. For $\sqrt{9/8}M<Q\leq(\sqrt{5}/2)M$, the
regions of stability are $r_{*}<r<r^-_{lsco}$ and $r>r^+_{lsco}$;
moreover, orbits in $r^-_{lsco}<r<r^+_{lsco}$ are unstable. This
situation is sketched and summarized in \fref{PcEqstab1m}(b).
Finally, for $Q>(\sqrt{5}/2)M$ all orbits with $r>r_*$ are stable.
Moreover, as $Q/M$ increases in the interval $[1,\sqrt{5}/2]$, the
energy and angular momentum decrease along $r=r^+_{lsco}$  and
increase along $r=r^-_{lsco}$.  For the classical radius $r=r_{*}$
the energy increases as $Q/M$ increases. This behavior is summarized
in \fref{PcEqstab1m}(c), see also \fref{PcEqstab1mtree}. A more
detailed discussion of these results will presented
elsewhere\cite{Mio1Que}  (see also Refs.~\refcite{Cohen:1979zzb} and
\refcite{Liang:1974ha}).
%
\section{Conclusion}
We discussed the motion of neutral test particles along circular
orbits in the Reissner-Nordstr\"om spacetime. We found that at the
classical radius $r=r_*= Q^2/M$ circular orbits exist with ``zero"
angular momentum. Inside the classical radius no time-like motion is
possible. The difference between black hole and naked singularity
configurations was studied in detail. In the case of black holes,
the radius of the last stable circular orbit has its maximum value
of $r_{lsco}=6M$ in the Schwarzschild limiting case, and reaches its
minimum value of $r_{lsco}=4M$  in the case of an extreme black hole
$Q=M$. In the case of naked singularities,  for a given value of the
ratio $Q/M$ two different values of $r_{lsco}$ are possible. As a
consequence, a non connected region of stability appears which can
exist only in configurations characterized by naked singularities.
%

\end{document}